\documentclass[seceq]{ptptex}

\usepackage{graphicx}

\title{
The Quest for Pionic and Kaonic Nuclear 
Bound Systems\\ Following Yukawa and Tomonaga%
}

\author{
Toshimitsu \textsc{Yamazaki}%
}

\inst{
Department of Physics, University of Tokyo, Bunkyu-ku, Tokyo 113-0033, Japan, and
RIKEN Nishina Center, Wako, Saitama 351-0198, Japan}

\abst{After sketching some historical events related to Yukawa and Tomonaga concerning the birth of mesons, the author describes recent developments in the spectroscopy of pion-nucleus bound states via ``pion-transfer" reactions. The role of pions as Nambu-Goldstone bosons in nuclear media is emphasized by recently obtained experimental evidence for the partial restoration of chiral symmetry breaking. New light is shed on $\bar{K}$ mesons, which play a unique role in forming dense nuclear systems. The basic unit, $K^-pp$, is predicted to possess a molecular structure with quasi-$\Lambda(1405)$ as an ``atomic constituent". We find here {\it super strong nuclear force} produced by a migrating real $\bar{K}$ meson in the Heitler-London-Heisenberg scheme in place of the normal nuclear force mediated by Yukawa's virtual mesons.}

\begin{document}

\maketitle

\begin{center}
 Invited talk at the Yukawa-Tomonaga Centennial Symposium, Kyoto,\\ December, 2006~~~
{\bf Prog. Theor. Phys.}, in press
\end{center}

\section{Introduction}

It is a great honor and pleasure to have an opportunity to give a talk in 
this historical Yukawa-Tomonaga Centennial Symposium. I would like to start (Section 2) by sketching some exciting 
 events related to the birth of the Yukawa meson, which are not well known even to Japanese physicists. The Yukawa theory was created and published in 1935 \cite{Yukawa:35}, following the work of Heisenberg in 1932 \cite{Heisenberg:32} on nuclear binding phenomena. Heisenberg was stuck on the idea of molecular binding applied to the nuclear force in terms of ``Platzwechsel" {\it a la} Heitler and London, who explained the H-H bonding in the hydrogen molecule quantum mechanically in 1927 \cite{Heitler:27}. Symbolically, we have
\begin{eqnarray}
{\it Molecular:~ Heitler-London-Heisenberg:}~~e^-p + p \leftrightarrow p + e^-p.
\end{eqnarray}\label{eq:Heitler-London}However, this idea was abandoned because this $e^-p$ could not be identified with the neutron, which had been discovered by Chadwick in 1931. 

In the spring of 1933, Yukawa was also struggling with the above idea of Heisenberg. Eventually, Yukawa hit on the idea of mediating virtual bosons, \cite{Yukawa:35} instead of {\it migrating real} electrons,
\begin{equation}
{\it Nuclear~ Force: Yukawa:}~~~p \leftrightarrow n +  \pi^+ , ~~~ n  \leftrightarrow p +  \pi^-.
\label{eq:Yukawa-vertex}
\end{equation}
\noindent
The name of $\pi$'s was given later after the discovery of  ``two mesons". 
 At that time Tomonaga was an assistant to Nishina, working on a    
theoretical explanation of new experimental data on the interaction between proton and  neutron, 
 employing various interaction forms. His results were communicated to Yukawa \cite{Tomonaga:33}, as we shall see shortly. Here, important interplay between Yukawa and Tomonaga emerged, and it continued in the years to come.  
 
In Section 3, I describe pion-nucleus bound systems, which have recently been produced experimentally using a particular ``synthesis reaction," named ``pion transfer reaction." I  emphasize that with the new understanding of pions as Nambu-Goldstone bosons, bound pions have come out to play an important role as a probe of chiral symmetry restoration, as revealed in recent experiments. In Section 4, I discuss nuclear bound states of $\bar{K}$, a strange relative of the Yukawa meson. This is a new field of nuclear physics, lying at the frontier of the study of {\it cold and dense} nuclear systems. Here, we present recent predictions that tightly bound dense nuclear systems may exist, even on non-existent nuclei, the most fundamental one being $K^-pp$. Experimental investigations of such phenomena are in progress. I emphasize at the end that the unsuccessful Heitler-London-Heisenberg scheme for nuclear binding, which was replaced by Yukawa's  scheme with a ``virtual mediating particle", has now been revived in nuclear $\bar{K}$ bound states as
\begin{equation}
{\it Super~Strong~ Nuclear~ Force:}~~~K^-p + p \leftrightarrow p + K^-p,
\end{equation}
where the {\it real migrating bosonic} particle $\bar{K}$ induces an enormous binding between nucleons.

\section{Historical events in the birth of mesons}\label{sec:historical}

\noindent
{\bf Yoshio Nishina: father of nuclear and particle physics in Japan }\\

Yoshio Nishina, well known after his famous work with Oscar Klein on the Klein-Nishina formula, played a very important role in the development of  modern science in Japan. He returned to Japan  
from Copenhagen, where he spent the years 1922--29 under Niels Bohr, and started a laboratory at RIKEN in 1929, with a spirit which he brought back from Copenhagen. This laboratory contributed greatly to the advancement of nuclear and cosmic-ray physics. His lecture on quantum physics at Kyoto University provided great stimulation to Yukawa and Tomonaga, who were then undergraduate students of physics. It is known that he advised to Yukawa to consider a boson instead of fermions as a mediating particle, with which Yukawa was struggling before the birth of the Yukawa theory \cite{Yukawa:35}. Their communications continued throughout Nishina's lifetime. Nishina adopted Tomonaga as an assistant in his laboratory at RIKEN. Later, Shoichi Sakata and Minoru Kobayashi joined this group. In addition to the cosmic ray research, Nishina constructed a Cockcroft-Walton accelerator, followed by a cyclotron, which produced exciting results. The most notable were the discovery of $^{237}$U \cite{Nishina:U237}, which would be the parent of a hitherto unknown $Z=93$ transuranium element, and the discovery of symmetric fission induced by fast neutrons \cite{Nishina:fission}. 

Nishina's idea of forming an open forum for inter-university collaboration among scientists, which he borrowed from Copenhagen, greatly affected Japanese physicists. Yukawa and Tomonaga were strongly influenced by Nishina, and contributed later to the birth of a new inter-university research institute in 1953, the Research Institute for Fundamental Physics in Kyoto (presently, the Yukawa Institute for Fundamental Physics). The second one, the Institute for Nuclear Study, University of Tokyo, was founded to develop nuclear physics, high-energy physics, cosmic-ray physics and theoretical physics. Tomonaga played a very important role in its creation. It is to be noted that these institutes were founded at the same time as CERN in a similar spirit. 

Nishina's communications with Yukawa and Tomonaga (and many others) in the 1930s and 1940s were recently published in three volumes  ``Collected Correspondence of Yoshio Nishina  \cite{Nishina-Letters}." A number of exciting events are revealed in these newly disclosed letters, to which I owe this talk.  \\

\begin{figure}[tbh]
\centering
\includegraphics[width=9cm]{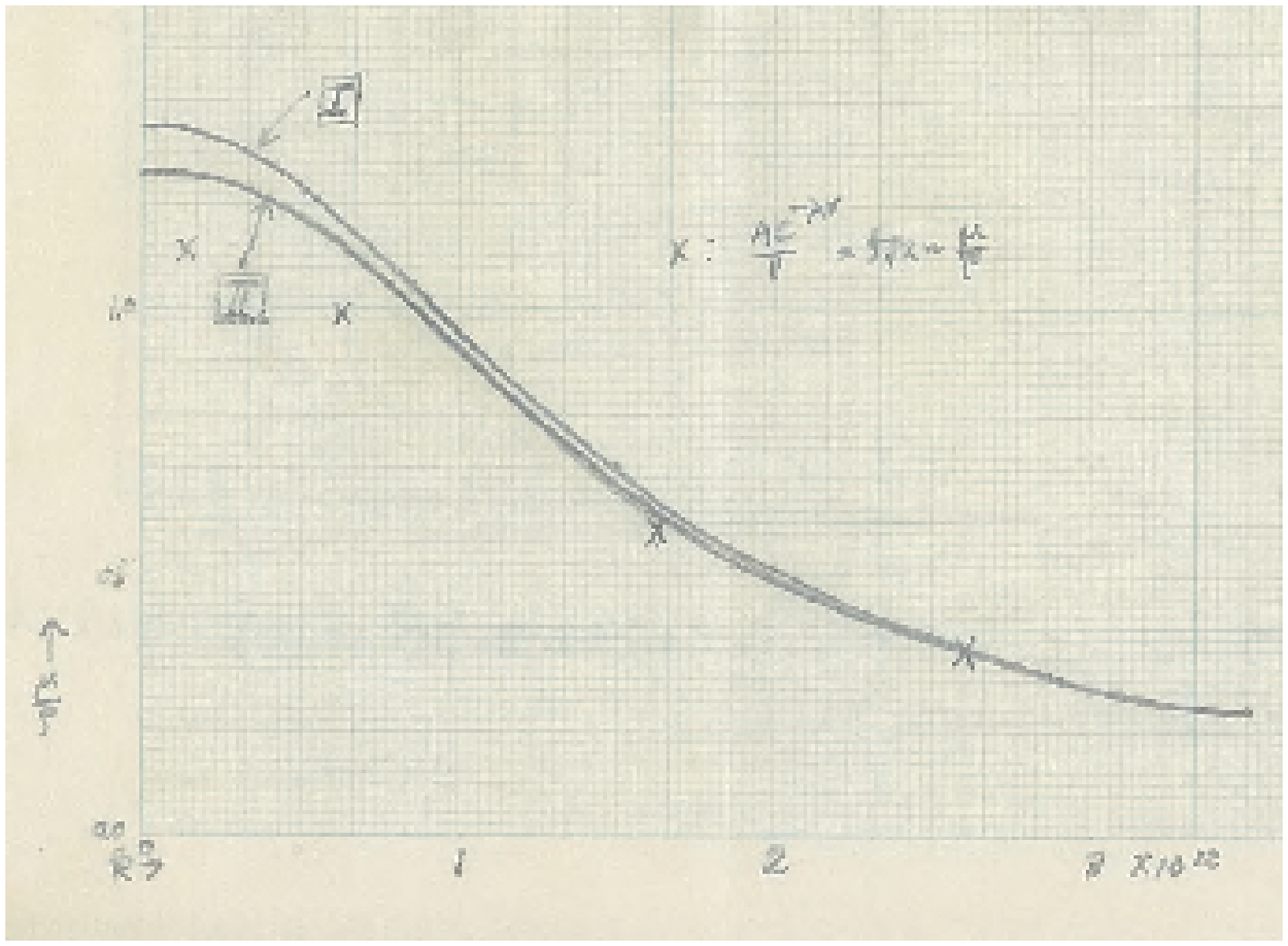}
\includegraphics[width=4.5cm]{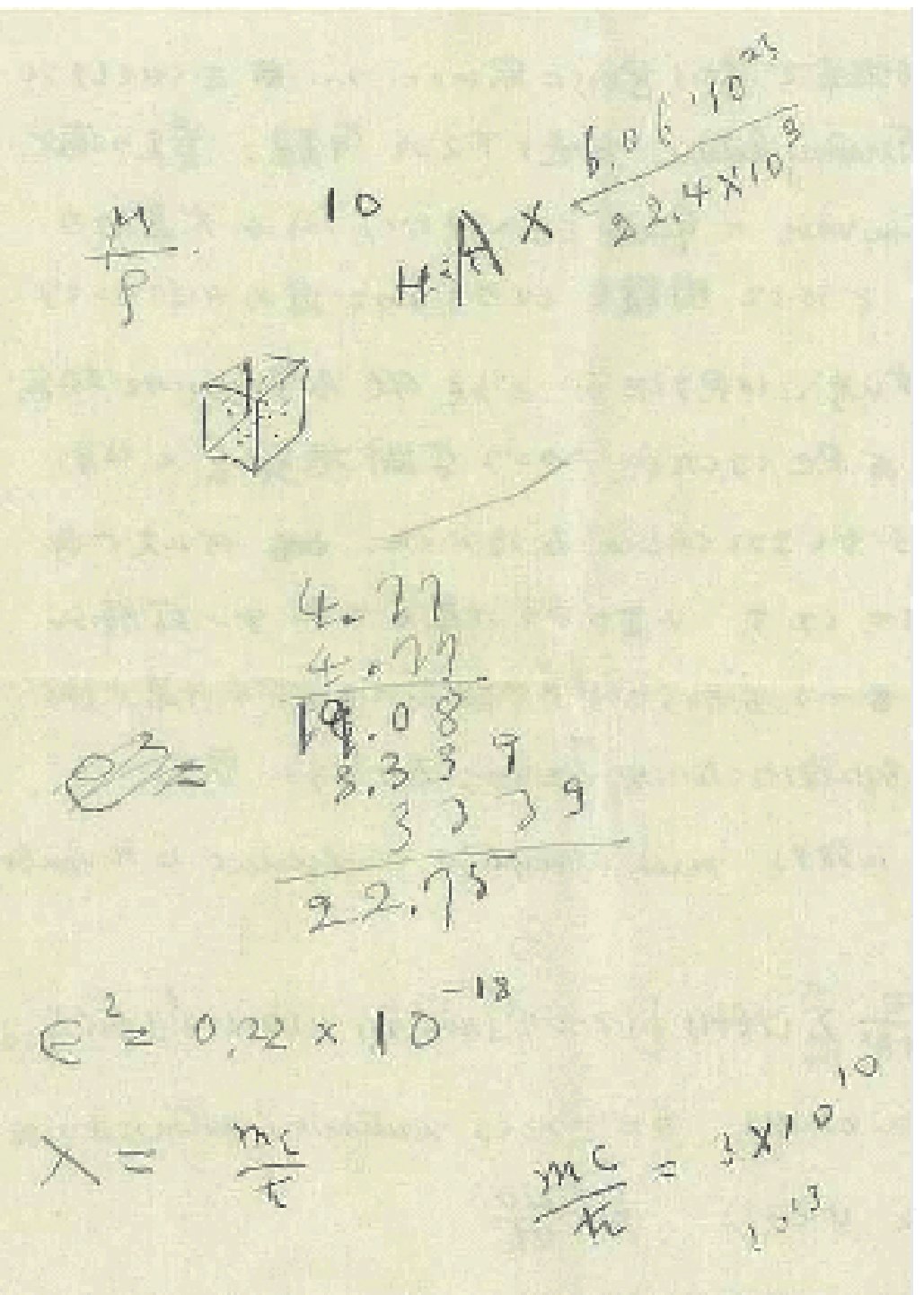}
\caption{(Left) Tomonaga's hand-drawn plot of the p-n reaction data with his theoretical fitting using the ``Yukawa interaction."  (Right) Yukawa's note on the back of Tomonaga's letter. From Ref. \cite{Tomonaga:33}.} 
\end{figure}\label{fig:Tomonaga-letter}

\noindent
{\bf Interplay between Yukawa and Tomonaga in the birth of mesons}\\

Right after the discovery of the neutron by Chadwick, the Nishina laboratory was studying the proton-neutron interaction. Tomonaga, as a resident  theorist, investigated the $p-n$ binding and $p-n$ reaction theoretically. To analyze and account for hot  experimental data, he employed various analytical forms for the p-n interaction. Tomonaga reported his results to Yukawa in a private letter \cite{Tomonaga:33} in the spring of 1933, shortly after Yukawa presented a talk on the problem of nuclear force at the Japan Phys. Math. Society meeting in Sendai. It is surprising that Tomonaga used the form, as shown in Fig.~\ref{fig:Tomonaga-letter} (left panel, a drawing attached in the letter), 
\begin{equation}
J(r) = A\, \frac{{\rm exp} (- \lambda r)}{r},
\end{equation}
for this interaction, as it is precisely  
 the form which was later called ``Yukawa interaction". At that time, however, neither of them seemed to recognize the deep meaning of this formula and, in particular, the meaning of the range parameter $\lambda$, which Tomonaga deduced from experimental data to be
\begin{equation}
\lambda = 7 \times 10^{12}~/{\rm cm}
\end{equation}

 It is extremely interesting that Yukawa jotted some notes in the back of this letter (see Fig.~\ref{fig:Tomonaga-letter}, right panel), such as  
\begin{equation}
\lambda_{Compton} = \frac{mc}{\hbar} \sim 3 \times 10^{10}~{\rm cm}^{-1}.
\end{equation} 
It is very interesting to speculate about what Yukawa was thinking when he made this hand-written calculation of the electron Compton wavelength. If we divide Tomonaga's value by this value, we would obtain the value 230 !! We can thus imagine that this letter must have had a profound influence on  Yukawa, who was in the midst of struggling with the problem of nuclear force in 1933, but had not yet formulated the idea of  the Yukawa interaction, in which the range parameter is related to the mass of the mediating particle $X$:
\begin{equation}
\lambda = \frac{m_X c}{\hbar}.
\end{equation}
 Tomonaga's work on the range of the p-n interaction was later mentioned in the footnote of Yukawa's first paper \cite{Yukawa:35}, whereas Tomonaga published this work only in 1936 \cite{Tomonaga:36}, 3 years after his letter to Yukawa. A similar work on the proton-neutron binding by Bethe and Peierls \cite{Bethe:35} appeared in literature in 1935.
\\

\noindent
{\bf Discovery of the ``mesotron"}\\

\begin{figure}[tbh]
\centering
\includegraphics[width=6.5cm]{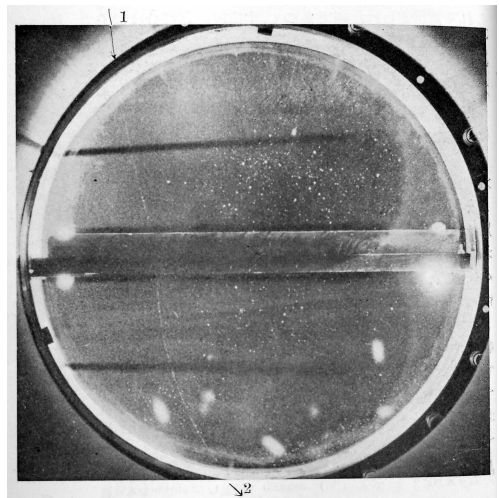}
\caption{Photograph of the cloud chamber track of a cosmic ray event taken by Nishina, Takeuchi and Ichimiya of RIKEN. From Nishina in ``Kagaku" \cite{Kagaku}.  }\label{fig:muon-track}
\end{figure}

Another story I would like to convey is that of the discovery of muons. Nishina constructed a large cloud chamber with a very strong and homogeneous magnetic field to measure cosmic rays. Around 1936-37,  there were four experimental groups in the world with the primary purpose of examining the validity of the Bethe-Heitler formula, which had just been derived. Neddermeyer and Anderson \cite{Nedd} were the first to report that there are some particles which do not obey this theory. Such particles were believed to be neither the proton nor the electron (positron), presumably having a mass between the proton and the electron. In the same year, similar findings were reported by other groups \cite{Jean}. Among them, two groups succeeded in the determination of the mass of such intermediate particles. The paper of  Nishina, Takeuchi and Ichimiya \cite{Nishina-muon}, reporting a value of $m_X/m_e = 180 \pm 20$, the most precise value at that time, was received by Physical Review on August 28, 1937, and was published on December 1. Interestingly, the paper of Steet and Stevenson \cite{Street}, reporting $m_X/m_e = 130 \pm 30$, was received on October 6, 1937, more than one month later than Nishina's paper, but was published on November 1, one month earlier. This situation resulted resulted from the fact that shipping 
 of the galley proofs back and forth took nearly 40 days. Figure \ref{fig:muon-track} shows a cloud chamber picture of Nishina' group, which was printed in a Japanese science journal ``Kagaku" in September 1937. 

Thus, it is fair to say that the two experiments were nearly of the same quality and significance. Nevertheless, the experiment of Nishina's group has hardly been recognized by physics community. It is a pity that even Japanese physicists are not aware of this great achievement. \\

\noindent
{\bf Precursors to great discoveries}\\

The experimental method of Nishina's group was to measure two tracks of a cosmic ray, before and after passing through a thick Pb absorber installed in the cloud chamber. In this way, they measured the energy loss $\Delta E$ versus the momentum $p$ for each event, and determined the mass of the particle. Surprisingly, Takeuchi reported in 1938 that he had found one event which exhibited a mass about half as large as the proton mass. One would guess it must have been the $K$ meson if this is correct, but Takeuchi 
admitted that he found only one such event. His collaborators and colleagues   
were very skeptical and discouraged him from publishing this result. For this reason he did not publish it,   but it was documented in a book \cite{Takeuchi}. This finding happened almost one decade prior to the discovery of the $K$ meson in 1947 by Leprince-Ringuet and L'heritier \cite{Leprince}. 
Many reminiscences of researchers at that time are collected in a book. \cite{Brown} 

At that time, it was not known that  the muon and the pion are distinct  
particles, and thus, the situation was very confusing. In 1937, far before the discovery of the pion, 
two female scientists at the Radium Institut in Vienna (presently, the Stefan Meyer Institute for Subatomic Physics), Marietta Blau and Hertha Wambacher, observed a star-like track
in emulsion \cite{Blau}, which may have been a pion-induced event. At that time, Tomonaga was at Heisenberg's Institute, and wrote a letter to Nishina, stating that Heisenberg was very absorbed by 
this kind of observation.  \\

\noindent
{\bf Tomonaga-Araki: gateway to exotic atoms}\\

Whereas the mass of the intermediate particle, called the ``mesotron" at that time, was accurately  determined in 1937, the nature of this particle remained a puzzle. It was difficult to reconcile the ``weak" nature of the mesotron with the Yukawa meson, a carrier of the strong interaction. There were many contradicting phenomena. In order to solve the decay and absorption puzzles of mesotrons, Tomonaga with Araki  theoretically studied the energy loss and capture processes of slow mesons \cite{Tomonaga-Araki}, clarifying that there is a marked difference between the nuclear capture probabilities of positive and negative mesons; i.e., the capture probability of a negative meson is of the order of $10^{12}$ s$^{-1}$, while positive mesons are not captured. However, this theory created a new puzzle, because, experimentally, negative mesons were captured in heavy nuclei, but decayed freely in light nuclear targets \cite{Conversi}. Thus, the validity of this theory came into serious doubt, 
 but later, with the discovery of the Yukawa meson ($\pi$) as a parent of the muon \cite{Powell}, all of these puzzles were solved. The Tomonaga-Araki paper was the first theoretical treatment of the formation of exotic atoms from negative mesons. Later, Fermi, Teller and Weisskopf \cite{Fermi} developed a comprehensive theory of the atomic capture of negative mesons.   \\

\noindent
{\bf From mediating persons to mediating virtual particles}\\

In Japanese, the Yukawa meson is written '†ŠÔŽq,  which literally means ``intermediate particle." Chinese people named  ‰îŽq, which means ``mediating 
particle." I find that this naming is very physically intuitive. It is also surprising that the character {\Large ‰î} looks very much like the Greek character {\LARGE $\pi$} (originating from ``primary") for the Yukawa meson. What an amazing accident !!

Yukawa's concept of a mediating virtual particle was strange to physicists at that time. It may be said that the Yukawa meson was born in accordance with the social custom in Kyoto. According to the customs in Kyoto, polite relation between two people is established through a direct interaction but, rather, through an indirect interaction communicated by a mediating person. In this sense, the old tradition of Kyoto became the origin of modern quantum physics.\\

\section{Pionic Nuclear Systems}

\noindent
{\bf Explicit roles of pions in nuclei}\\

The Yukawa meson was introduced as a mediating virtual particle for the nuclear force, as expressed in (\ref{eq:Yukawa-vertex}). 
Although the pion plays an essential role in nuclear binding, it does not play an explicit role in nuclear physics. The pion does not appear as a real particle, but is hidden as a virtual particle. However, there are some interesting exceptions. The first one concerns the role of the 
pion-exchange current, which produces an extra orbital magnetic moment of a nucleon. This was first pointed out by Miyazawa in 1951 \cite{Miyazawa:51}, who showed theoretically that the orbital $g$-factor of the proton [$g_l (p)$]  
increases by approximately 0.1 and that of the neutron [$g_l (n)$] decreases to $-0.1$. Clear experimental  evidence was obtained in 1970 from the measurement of the magnetic moment of the $11^-$ state \cite{Yamazaki:70},  which possesses an orbital angular momentum of  $11 \hbar$, with an almost vanishing intrinsic spin contribution. Around this time Riska and Brown \cite{Riska} found that the known anomalous photo disintegration rate of the deuteron (or, equivalently, the rate of the p-n capture reaction yielding the deuteron, which was Tomonaga's first work  in 1933) was accounted for by taking into account the meson exchange current. Then, Fujita and Hirata \cite{Fujita-Hirata} showed that the enhancement of the orbital $g$-factor is related to the enhancement in the giant dipole 
resonance in nuclear phenomena. More detailed accounts are given elsewhere 
\cite{Yamazaki:79,Rho-Wilkinson}. Later, I deduced the effective nuclear magneton by combining $g_l (p)$ and $g_l (n)$ values, which was found to be enhanced by about 8 \% \cite{Yamazaki:85}. \\ 

\noindent
{\bf Syntheses of exotic atoms and nuclei }\\

Now I have come to the main subject of my talk, namely, the synthesis of quasi-stable exotic matter 
with pions and kaons as constituents. This seems to be very 
difficult because hadrons, which are strongly interacting particles, are 
strongly absorbed in nuclei, and thus are very short-lived constituents of 
matter.  There are three recently discovered exceptions. The first is an anti-protonic helium atom, which is an atom-molecular state. The second is deeply bound 
pionic nuclear states, with an excitation of 140 MeV from the nuclear ground state. The third is possible kaonic cluster 
nuclei. All of these 
can be regarded as kinds of Feshbach resonances, because they are 
bound states from the viewpoint of some particles, but they are embedded in a continuum. These three different species have different mechanisms for quasi stabilities. \\

\noindent
{\bf Antiprotonic helium}\\

Tomonaga and Araki treated the Coulomb-capture effect, but the primordial 
bound state in which this negative particle is captured by the nucleus was 
very much in a mist.  Later, Fermi and Teller \cite{Fermi} clarified  
that the atomic quantum numbers of the primordial states are close to $n = \sqrt{M_X/m_e}$. In the case of a pion, it is about 14, and in the case 
of an antiproton, it is about 40. In general, it is very large, but it is almost 
impossible to identify these individual states.  
There is one exception, that of the metastable antiprotonic helium atom.  This form of matter was discovered in 1991 at KEK in Japan \cite{Iwasaki:91};  
3 \% of the antiprotons implanted into liquid helium underwent a very delayed 
annihilation. This puzzling phenomenon was later understood fully \cite{Yamazaki:PhysRep}.  This atom   is a miraculously 
long-lived interface between the matter world and the anti-matter 
world, and laser spectroscopy has been developed to the degree to test the CPT symmetry between the proton and the antiproton to high precision.  \\

\noindent
{\bf From virtual pions to real pions in nuclei}\\

Pionic bound states have been studied as pionic atoms. The year 1966 marked several breakthroughs in pionic atom spectroscopy. The first was the theoretical development of the pion-nucleus optical potential in the form of the Ericson-Ericson theory \cite{Ericson:66}. 
The potential parameters in the optical potential can be deduced by combining pionic-atom x-ray data of the strong-interaction shifts (the deviation from the pure Coulomb case) and widths.

\begin{figure}[h]
\begin{center}
\includegraphics[height=6.5cm]{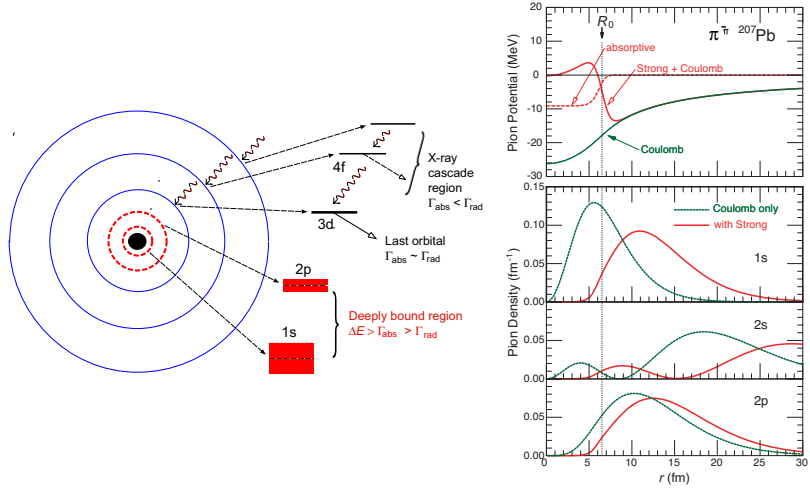}
\caption{
(Left) A schemetic figure of pionic atom states with X-ray
transitions down to the last orbital and deeply bound inner orbits with
large widths which cannot be populated following the x-ray cascade. They
have large widths due to nuclear absorption but they are still discrete states 
with $\Gamma_n < E_n - E_{n-1}$.
(Right) Mechanism for the narrow pionic bound states. Taken from Toki {\it et al.} \cite{Toki:88}.
The upper figure shows the pion optical potential for
$^{208}$Pb.  The finite-size Coulomb potential is expressed by the dotted curve, 
and that with the optical potential by the solid curve. The imaginary
part is depicted by the dashed curve.  The lower figure shows
the pionic wavefunctions of the 1s, 2s and 2p states in coordinate
space.  The dashed curves and the solid curves were obtained with
a finite-size Coulomb potential and with an optical potenial.
The half-density radius $R_0$ of $^{208}$Pb is indicated by the
broken line.
}
\label{fig:deeply-bound}
\end{center}
\end{figure}

The first precise measurement of the strong-interaction shift and width was made in pionic atoms at the historical 184-inch cyclotron of Berkeley \cite{Jenkins}. Since then,  a great amount of experimental data, not only on pionic atoms but also on other exotic atoms, has been accumulated, as reviewed by Batty, Friedman and Gal \cite{Batty:93}. The systematics of the strong-interaction shift in the pionic 1s states in light nuclei indicate that the s-wave part of the optical potential is repulsive. The shift increases rapidly with the atomic number $Z$, but the $2p \rightarrow 1s$ transition vanishes beyond $Z \sim 28$, due to the increasing nuclear absorption in the $2p$ state. Pionic x-ray transitions in heavier nuclei are measurable only for higher orbitals, such as 2p, 3d, and so on, as schematically shown in Fig.~\ref{fig:deeply-bound}. Then, we are led to the question of whether or not the 1s ground states in heavy nuclei exist. If so, how can we reach it? This is the question that Toki and I posed and answered in 1988 \cite{Toki:88}. \\

\noindent
{\bf Deeply bound pionic states}\\

There was the prevailing belief at that time that there was no bound state beyond (namely, inside of) the last orbital, where the pionic x-ray cascade terminates due to nuclear absorption. However, this erroneous belief does not make sense, because whether a 
state exists or not and whether a state is populated by some path are 
two different things. Namely, the existence of a state has nothing to do with the population of that state. The criterion for a state to ``exist" is the {\it discreteness of the state, $\Gamma < \Delta E$, namely, that the particle undergoes orbiting motion within its lifetime}. 
  
What does one expect for the width? A naive estimate is $\Gamma \sim 2 W_0 \sim 20$ MeV, considering the imaginary potential, $W_0 \sim 10$ MeV, deduced empirically. This estimate would make sense if the bound pion were to reside fully inside the nucleus.  However, the $\pi^-$ meson is pushed away from the nucleus by the repulsive interaction and overlaps with the nucleus only partially, as shown in Fig.~\ref{fig:deeply-bound}. This ``narrowing mechanism" was first pointed out by Freidman and Soff \cite{Friedman-Soff} in 1985. Generally, deeply bound states of any negative hadron have such a discrete character, as Coulomb-assisted hybrid bound states \cite{Yamazaki:88}. Toki and collaborators \cite{Toki:88} showed that all the pionic bound states in heavy nuclei are discrete and thus can be accessed with some experimental method, that is, the ``pion-transfer reaction".  

 Another interesting 
aspect of deeply bound pionic states is that the $1s$ bound state energy of a bosonic atom, as given by the Klein-Gordon equation, decreases  with the increase of $Z$ to the point where the total energy becomes zero (at $Z = 137/2$ for a poin nucleus, in contrast to the fermion case of $Z = 137$). Here, the binding energy is as large as the pion rest mass. We can conceive of a pionic atom with $Z  > 137/2$, though in 
reality this interesting situation cannot be realized, because of the finite nuclear size, which causes the amount by which the energy decreases to be smaller.  \\

\begin{figure}[hb]
\begin{center}
\includegraphics[scale=0.38]{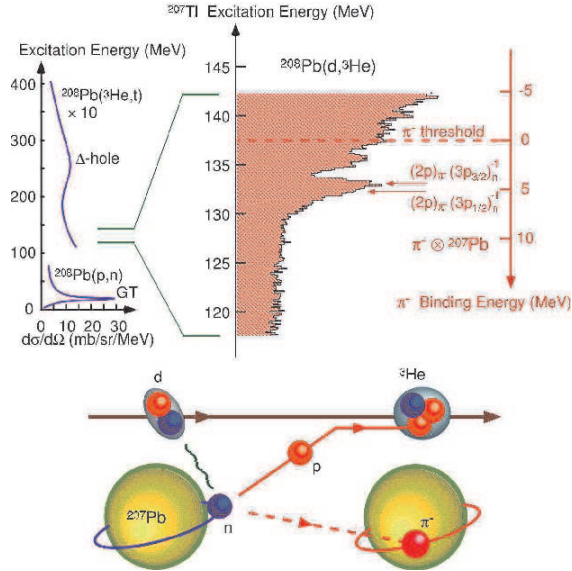}
\vspace{0.5cm}
\caption{(Upper) The observed $^{208}$Pb ($d,{\rm ^3He}$) reaction spectrum revealing the predicted deeply bound $\pi^-$ states. (Lower) The mechanism to produce $\pi^-$ states from ``inside".}
\label{fig:GSI}
\end{center}
\end{figure}

\noindent
{\bf Pion transfer reactions: the first success}\\

Toki and collaborators \cite{Toki:88,Toki:91} proposed ``pion transfer" reactions to populate pionic bound states in which a real pion is bound. This has an 
interesting name in honor of Yukawa, because the reaction vertex is merely the Yukawa vertex, Eq.(\ref{eq:Yukawa-vertex}). 
Whereas the Yukawa vertex itself cannot be observed in free space, because it does not satisfy the conservation of energy and momentum, the pion produced in a nucleus can form a pionic bound state.    We noticed that a 
$(d,^{3}$He) reaction would be a very suitable one, as the momentum transfer is small \cite{Toki:91}.  
We carried out the experiment using the 600 MeV deuteron beam from the SIS18 accelerator of GSI, and succeeded in observing the spectrum with the $2p$ and $1s$ bound states of $\pi^-$ in $^{207}$Pb  \cite{Yamazaki:96}, as shown in Fig.~\ref{fig:GSI}. 
This experimental spectrum is in good agreement with the predicted one \cite{Toki:91}. We can say that this agreement provides experimental verification of the Yukawa vertex.  
\\

\noindent
{\bf Pions as Nambu-Goldstone bosons}\\

Recently, Weise \cite{Weise:00} and Kienle and Yamazaki \cite{Kienle:01,Kienle:04} have pointed out that pionic bound states in nuclei could be a unique indicator of chiral symmetry restoration in nuclear medium. The quark condensate $ \langle \bar{q} q \rangle$ in the QCD vacuum introduced by the spontaneous breaking of chiral symmetry is believed to be the origin of the large hadron masses ($\sim 1$ GeV), as compared with the very small masses (several MeV) of their constituents, the $u$ and $d$ quarks \cite{Nambu}. This situation can be examined by applying some external force, just as superconductivity can be studied by applying an external magnetic field. In this framework, the quark condensate in nuclear media is expected to decrease with the increase of the nuclear density and the temperature, as studied by Hatsuda and Kunihiro \cite{Hatsuda85} and Vogl and Weise \cite{Vogl91}. This may be observed as changes of hadron masses in nuclear media \cite{Brown-Rho}. However, the ``invariant-mass spectroscopy applied to decaying hadrons in nuclear media has inherent difficulties, as clarified in Ref. \cite{Yamazaki:99}. A new strategy is to study how the isovector $s$-wave interaction, represented by the parameter $b_1$ in Eq.(\ref{eq:b1}), changes in nuclei. This approach is outlined below. \\

 \noindent
{\bf Pionic bound states as a probe of chiral symmetry restoration}\\

Hadrons are quasi-particle excitations of this vacuum separated by
an energy gap ($\Delta \approx 1$ GeV), whereas
pions are Nambu-Goldstone bosons of the vacuum state.
According to the low-energy theorem of Tomozawa and Weinberg \cite{Tomozawa,Weinberg}, the isovector s-wave part of the $\pi N$ interaction (represented by $b_1$) is connected to the pion decay constant, $f_{\pi}$ (=92.4 MeV), as follows:
\begin {equation}
T^{(-)} = \frac {1}{2} (T_{\pi ^- p} - T _ {\pi ^- n} ) \equiv - 4\pi,
\varepsilon _ 1 b_1 = \frac{m_{\pi}}{2f_\pi^2}
\label{eq:TW}
\end {equation}
with $\varepsilon_1 =   1 + m_\pi/M = 1.149$. The parameter $b_1$ in the optical potential is the most important one in this context. 
The quantity $f_{\pi}^2$ is the order parameter of 
chiral symmetry breaking and is related to the quark condensate
through the Gell-Mann-Oakes-Renner relation \cite{GOR}, 
\begin {equation}
m_\pi^2 f_\pi ^2 = -m_q  \langle \bar{u}u+\bar{d}d \rangle _0,
\label{eq:GOR}
\end {equation}
which yields $\langle \bar{q}q \rangle _0 \approx -(250~{\rm MeV})^{3}$ in vacuum. When a $\pi^-$ is implanted in a nuclear medium of density $\rho$,  a new vacuum state with a reduced condensate,
$\langle \bar{q}q \rangle _{\rho}$, is  created \cite{Hatsuda85}.
The density dependence of the quark condensate is expressed to the
leading order in the pion-nucleon $\sigma$-term
($\sigma_N \approx 45~{\rm MeV}$) in the form \cite{Drukarev}
\begin{equation}\label{eq:condensate}
       \frac{\langle \bar{q}q
\rangle _{\rho}}{\langle \bar{q}q
\rangle _{0}} \approx 1 - \frac{\sigma_N}{m^2_\pi f_{\pi}^2}\rho,
\end{equation}
which yields a reduction factor of about 0.65 for normal nuclear density,
$\rho = \rho_0 = 0.17~{\rm fm}^{-3}$.
Similarly, the pion decay constant in a medium
(identified as the  time component of the
axial current) is reduced as~\cite{Thorsson}
\begin{equation}\label{eq:fpi}
\frac{f_{\pi}^* (\rho)^2}{f_{\pi}^2} \approx 1 - \alpha \rho,
\end{equation}
where the parameter $\alpha$ is predicted to be $\alpha \rho_0 \approx
0.45$ from the chiral dynamics \cite{Meissner2002}. This reduced pion decay
constant is associated with the in-medium isovector $\pi N$ scattering
length as \cite{Kolomeitsev:02}
\begin{equation}\label{eq:b1}
R(\rho) = \frac{b_1^{\rm free}}{b_1^* (\rho)} \approx \frac{f_{\pi}^*
(\rho)^{2}}{f_{\pi}^{2}}.
\end{equation}
\\

\noindent
{\bf Partial restoration of chiral symmetry revealed}\\

\begin{figure}[tbh]
\centering
\includegraphics[width=7cm]{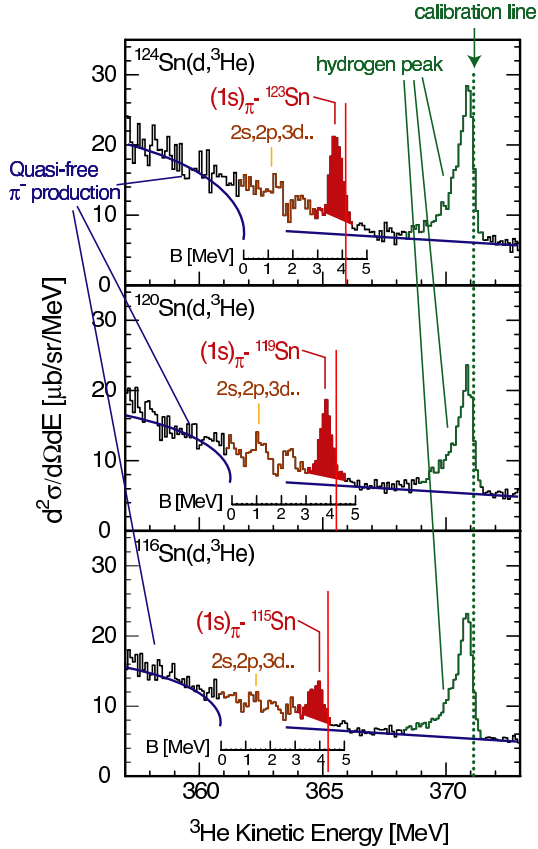}
\caption{Experimental spectra of the reactions $^{116,120,124}$Sn($d,^3$He) \cite{Suzuki:04}. The expected positions without the chiral symmetry restoration effect are indicated by vertical red lines, from which the observed 1s peaks (red) deviate significantly.}\label{fig:Sn-spectra}
\end{figure}

A dedicated experiment was planned and carried out at GSI to obtain the isovector s-wave interaction. For this purpose, the experimental data from 1s states of $\pi^-$ in heavy ($N \gg Z$) nuclei are essential. 
To produce the 1s $\pi^-$ states dominantly, we chose Sn 
isotopes as targets for $(d,^3$He) reactions with the ``recoilless condition" ($T_d$ = 500 MeV), and a precise spectroscopic investigation was carried out \cite{Suzuki:04}. We 
observed the 1s $\pi^-$ state in each of the  three isotopes, as shown in Fig.~\ref{fig:Sn-spectra}. The data are in good agreement with theoretical spectra \cite{Hirenzaki}. 

The magnitude of $|b_1|$, as deduced from this experiment, is found to be significantly greater than
 the free $\pi N$ value, which implies a reduction of 
${f_{\pi}^*}^2$ as 
\begin{equation}
R(\rho_e)  = \frac{b_1^{\rm free}}{b_1^* (\rho_e)} =  0.78 \pm 0.05.
\end{equation}
Since the bound $\pi^-$ probes the optical potential at an effective 
nuclear density $\rho_e  \approx
0.60\,\rho_0$ \cite{YH02}, the above value implies that the chiral order
parameter, $f_{\pi}^* (\rho_0)^2$, would be reduced by a factor of $0.64
\pm 0.08$ if the $\pi^-$ were embedded in the center of the
nucleus. Using Eq.(\ref{eq:fpi}) in the analysis, we
obtain an experimental value of $\alpha \rho_0 \approx 0.36 \pm 0.08$,
which is close to the value 0.45 predicted by chiral perturbation
theory~\cite{Meissner2002}.
If the theoretical value, $m_{\pi}^* \approx m_{\pi} + 3~{\rm MeV}$
(averaged over $\pi^+$ and $\pi^-$ \cite{Meissner2002}), is inserted
into the in-medium Gell-Mann-Oakes-Renner relation
\cite{Hatsuda85,Thorsson}, $\langle \bar{q}q
\rangle _{\rho_0}/\langle \bar{q}q
\rangle _{0}$  will be $(m_{\pi}^*/m_{\pi})^2 \times (1 - \alpha \rho_0)
\approx 0.67 \pm 0.08$,
which is in good agreement with the value 0.65, obtained from 
Eq.(\ref{eq:condensate}). Thus, clear evidence for the partial 
restoration of chiral symmetry is obtained from {\it well-defined} pionic states 
in a {\it well-defined} nuclear density.
 

\section{Kaonic Nuclear Systems}

\noindent
{\bf From pions to kaons: another Nambu-Goldstone boson}\\

Now I move to the final topic, that is, kaonic nuclear bound states. This is an extremely interesting but  still controversial subject.  
In free space, the bare $\bar{K}$-$N$ interaction in its 
$I=0$ channel is strongly attractive, as evidenced by the $\Lambda(1405)$ resonance (hereafter called $\Lambda^*$), which is largely populated in the $K^-$ absorption at rest in $^4$He \cite{Riley:75} and also in nuclear emulsions \cite{Davis:77}. 
Theoretically, both meson exchange \cite{Mueller:90} and chiral dynamics \cite{Weise:96}  treatments predict a strong attraction. When $\bar{K}$ is in a nuclear 
medium, the situation is not clear.  One view is that $\bar{K}$  
in nuclear matter of infinite extent, namely in continuum and scattering 
states, attenuates this strong attraction. This was first pointed 
out by Lutz \cite{Lutz}, and it has been used by many people to claim that the 
$\bar{K}$-nucleus interaction should be weak. This argument, however, is {\it valid only for unbound continuum states of infinite matter, and it cannot be applied to discrete bound states}. Akaishi and Yamazaki \cite{Akaishi:02}, after deriving bare $\bar{K}N$ interactions from a coupled-channel treatment   , so as to be fully consistent with the empirical observables (the energy and width of $\Lambda(1405)$ and low energy scattering lengths), treated few-nucleon systems with a $\bar{K}$. They predicted discrete bound states of $\bar{K}$, where the strong bare interaction persists. Furthermore, they showed that the nucleus can be shrunk by this strong interaction to a density 2-3 times larger than the normal nuclear density, $\rho_0 \sim 0.17$ fm$^{-3}$. \\

\noindent
{\bf $K^-pp$, the basic unit of $\bar{K}$ bound systems}\\

The simplest and most basic $\bar{K}$ nuclear system is $K^-pp$, first predicted in Ref. \cite{Yamazaki:02}. It has a large binding energy, 48 MeV, though the two protons are unbound without $K^-$. We have carried out comprehensive few-body calculations and clarified many interesting features. as summarized below.

\begin{figure}[htb]
\centering
\includegraphics[width=12cm]{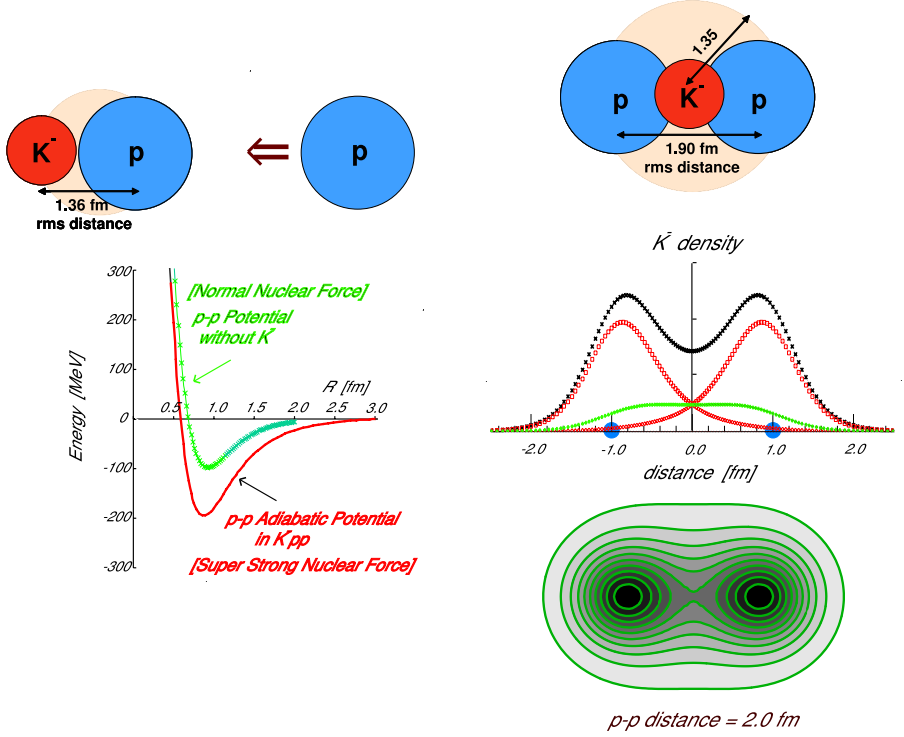}
\vspace{0cm}
\caption{\label{fig:AdiaPot} 
(Left) The adiabatic potential in the case that a proton approaches a $\Lambda(1405)$ of the form of  $K^-p$ as a function of the distance between $p$ and $p$. The Tamagaki potential $V_{NN}$ is shown for comparison.
(Right) The molecular structure of $K^-pp$. The contour (lower) and projected (middle) density distributions of $K^-$ in $K^-pp$ with a fixed $p-p$ distance (= 2.0 fm). }
\end{figure}

The kaonic nuclear cluster $K^-pp$ can be interpreted as a kaonic hydrogen molecule in the sense that $K^-$ migrates between the two protons, producing ``strong covalency" through the strongly attractive $\bar{K}N^{I=0}$ interaction. This is essentially the mechanism of Heitler and London \cite{Heitler:27} for the hydrogen molecule, though the nature of the interaction is completely different, and the mass of the migrating particle is much heavier and bosonic. This aspect is more clearly seen when the density distribution is plotted, with a fixed axis of the two protons. Figure~\ref{fig:AdiaPot} (left panel) shows the adiabatic potential, in the case that a proton approaches a $\Lambda(1405)$ particle, as a function of the $p$-$p$ distance. The deep potential indicates that a proton approaching an isolated $\Lambda^*$ from a large distance quickly becomes trapped and dissolved into the bound state of $K^-pp$. This leads to a $\Lambda^* p$ doorway situation following the $\Lambda^*$ doorway. Figure~\ref{fig:AdiaPot} (right panel) shows the projected distribution of $K^-$ along the $p$-$p$ axis and the contour distribition of $K^-$ in the case that the $p$-$p$ distance is fixed to 2.0 fm. (This case resembles the ground state of $K^-pp$, as the calculated rms distance is 1.9 fm.) The $K^-$ is distributed not around the center of $p$-$p$, but around the two protons. The $K^-$ distribution is composed of the ``atomic" part, as shown by the red dotted curves, and the exchange part (green broken curve) {\it a la} Heitler and London. 

We emphasize that the strong $I=0$ $\bar{K}N$ attraction produces a large exchange integral,
\begin{equation}
\langle v_{\bar{K}N} (12) + v_{\bar{K}N} (13) \rangle_{ab+ba} = -52.6~{\rm MeV},
\end{equation}
which is the source for the deeper binding of $K^-pp$ in comparison with $K^-p$.
Despite the drastic dynamical change of the system caused by the strong $\bar{K} N$ interaction, the identity of the ``constituent atom," $\Lambda^*$, is nearly preserved, because of the presence of a short-range repulsion between the two protons. The molecule $K^-pp$ is a tightly bound $\Lambda^*-p$, which we call the $\Lambda^* p$ doorway in the formation process. \\

\begin{figure}[h]
\centering
\includegraphics[width=14cm]{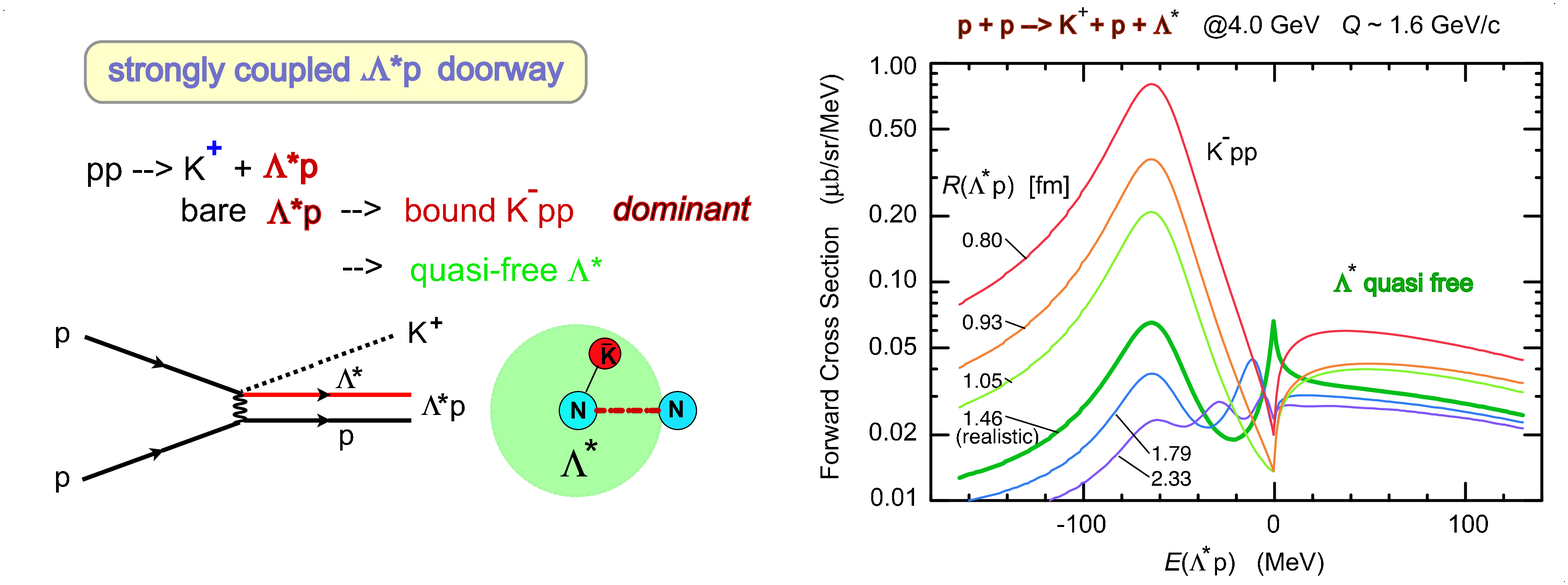}
\vspace{0cm}
\caption{\label{fig:pp2KX} 
 (Left) Diagram for the $p(p,K^+)K^-pp$ reaction. (Right) Calculated spectral shape for several rms distances $R(\Lambda^* p)$, arbitrarily chosen.
}
\end{figure}

\noindent
{\bf $K^-pp$ production in $NN$ collisions}\\

Now we consider the following process with a projectile proton and a target proton:
\begin{equation}
p + p          \rightarrow K^+ + (\Lambda^* p)
         \rightarrow  K^+ + K^-pp. \label{eq:pp2KLp},
\end{equation}
 where the $\Lambda^* p$ doorway state proceeds to $K^-pp$. The formed $K^-pp$ decays not only via the major channel, 
$K^-pp \rightarrow \Sigma + \pi + p$,
but also through non-pionic decay channels. The reaction diagram is shown in Fig.~\ref{fig:pp2KX} (left pannel). This problem is discussed fully in Ref. \cite{Yamazaki:06a,Yamazaki:07a}.

The $p \rightarrow p + K^- + K^+$ process, in which a $K^+K^-$ pair is assumed to be created at zero range from a proton, is strongly off the energy shell ($\Delta E \sim 2 m_K$). This process is realized only with a large momentum transfer to the second proton, and it is efficient, due to by a short-range $pp$ interaction. 

The calculated spectral function for $T_p$ = 4 GeV at a forward angle on the scale of $E(\Lambda^* p) = 27~{\rm MeV} - B_K$ is presented in Fig.~\ref{fig:pp2KX} (right pannel). Surprisingly, in great contrast to ordinary reactions, the spectral function is peaked at the bound state with only a small quasi-free component. This means that the sticking of $\Lambda^*$ and $p$ is extraordinarily large.
This dominance of $\Lambda^* p$ sticking in such a large-$Q$ reaction can be understood as originating from the matching of the small impact parameter with the small size of the bound state. 
It is vitally important to examine our results experimentally. An experimental observation of $K^- pp$ in  a $pp$ collision will not only confirm the existence of $K^- pp$ but also prove the compactness of the $\bar{K}$ cluster.  

\begin{figure}[h]
\centering
\includegraphics[width=14cm]{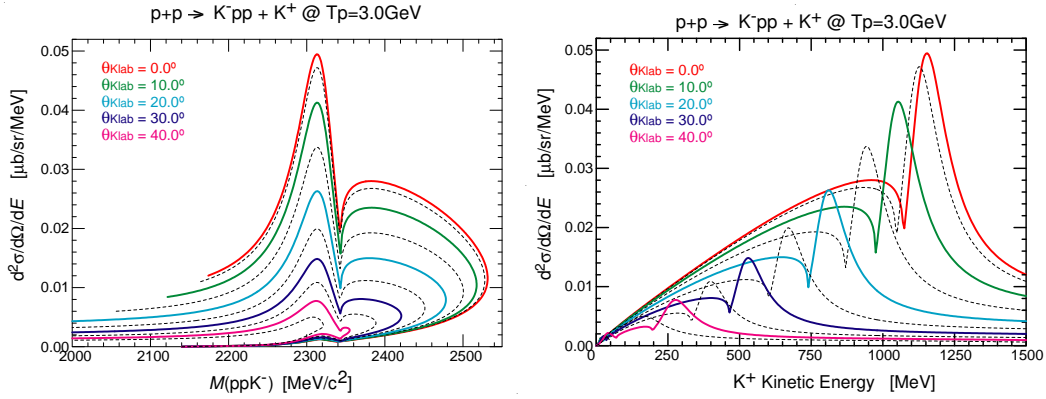}
\vspace{0cm}
\caption{\label{fig:Cross-section} 
Predicted differential cross sections of $p + p \rightarrow K^+ + X$ at $T_p = 3.0$ GeV. (Left) $M_X$ spectra at various $K^+$ laboratory angles. (Right) $K^+$ energy spectra at various $K^+$ laboratory angles. }
\end{figure}

The reaction we propose is essentially a reaction of two-body final states,  
$p + p \rightarrow K^+ + X$,
where the unknown object $X$ with a mass $M_X$ can be searched for in a missing mass spectrum of $K^+$, $MM(K^+)$. The calculated cross sections for $T_p$ = 3 GeV at various laboratory angles for an assumed bound-state mass of $M_{K^-pp}$ = 2310 MeV/$c^2$ are presented in Fig.~\ref{fig:Cross-section}.  

The elementary reaction of the type 
\begin{equation}
p + p \rightarrow K^+ + Y^0 + p
\end{equation}
was studied at an incident proton energy of $T_p$ = 2.85 GeV by the DISTO group at SATURNE \cite{DISTO}, which identified $\Lambda$ from  the invariant-mass spectrum of $p + \pi^-$. The most important information within the present context is contained in the spectrum of $MM(K^+)$, which is related to the mass of $K^- pp$, 
but no such spectrum has been reported. 
Now, a new experiment of the FOPI group at GSI \cite{FOPI-proposal} is in progress. Its aim is to measure all of the products in the $p + p$ reaction at $T_p$ = 3 GeV to reconstruct both the invariant mass $M_{inv} (\Lambda p)$ and $MM(K^+)$. \\

\noindent
{\bf $\bar{K}$-induced shrunk nuclei}\\

\begin{figure}[hb]
\begin{center}
\begin{minipage}[t]{0.35\textwidth} 
\includegraphics[width=1.00\textwidth]{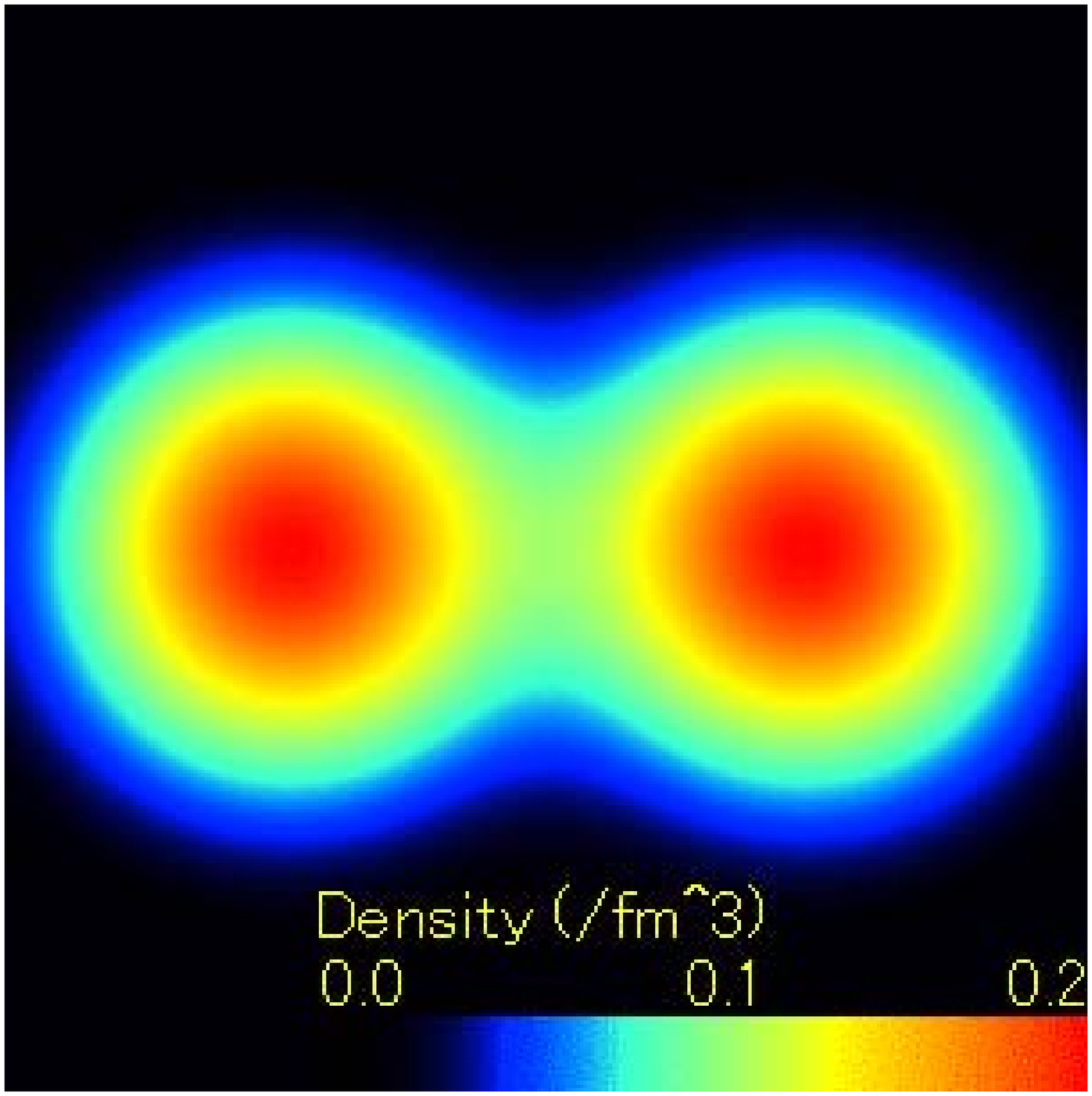}%

(a) $^8$Be
\end{minipage}
\hspace{0.3cm}
\begin{minipage}[t]{0.35\textwidth}
\includegraphics[width=1.00\textwidth]{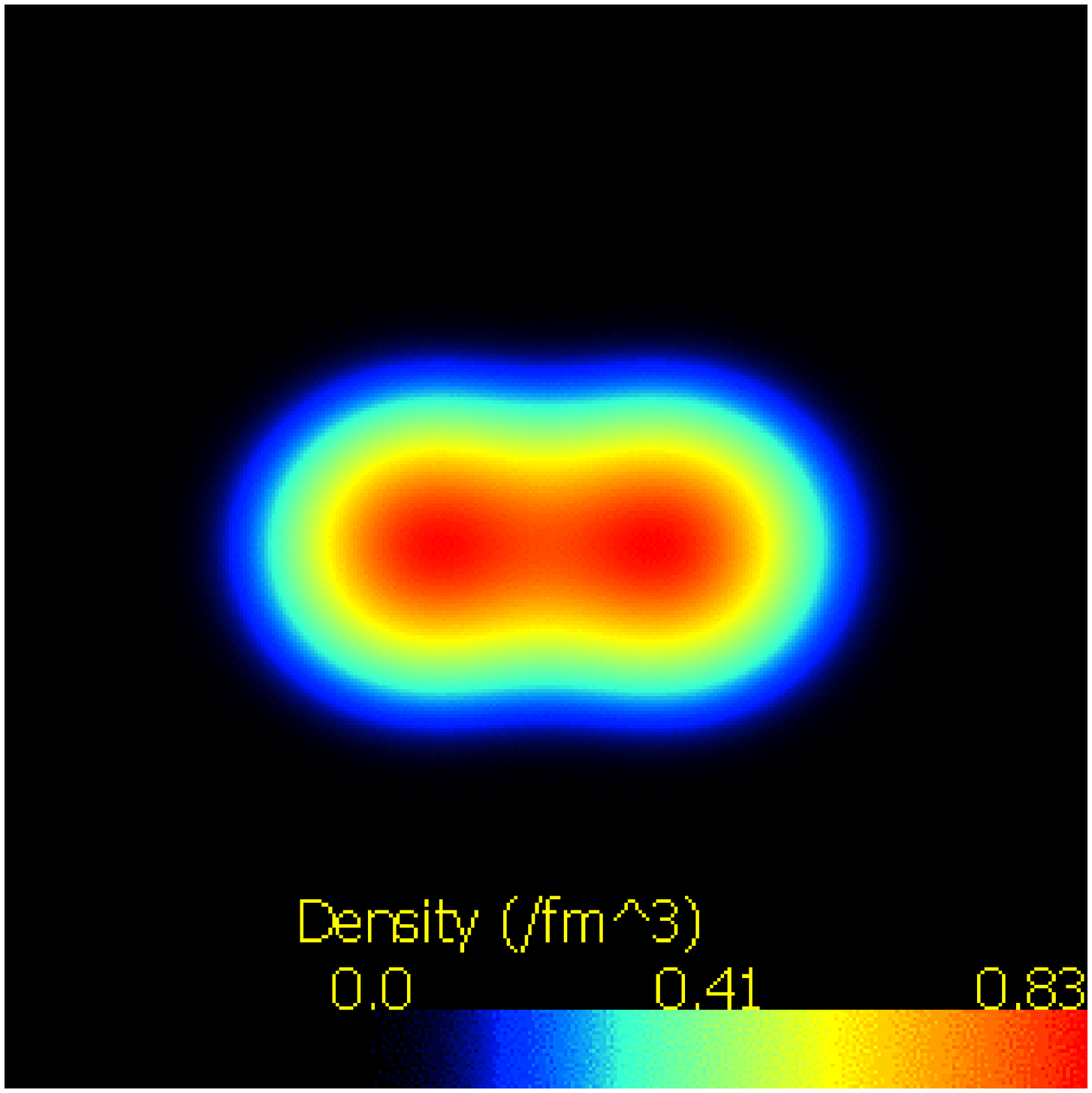}%

(b) $^8$BeK$^-$
\end{minipage}

\caption{ \label{fig:BeK}  Comparison of the calculated density contours of (a) ordinary $^8$Be and
(b) $^8$BeK$^-$. Each figure depicts a size of  7 $\times$ 7 fm. From Dot$\acute{\rm e}$ {\it et al.} \protect\cite{Dote:04a,Dote:04b}.  
}
\end{center}
\end{figure}

Since the $I=0$ $\bar{K} N$ interaction is strongly attractive without a hard core of short range, a $\bar{K}$ can form very unusual condensed nuclear systems, which do not exist in nature, as shown in Refs.~\cite{Akaishi:02,Yamazaki:02,Dote:04a,Dote:04b,Yamazaki:04,Akaishi:05a}. For example, Fig.~\ref{fig:BeK}, depicting a result of an antisymmetrized molecular dynamics calculation 
\cite{Dote:04a,Dote:04b}, shows dramatically how the $^8$Be nucleus, which is composed of two alpha clusters, is shrunk by adding a $K^-$ meson. It is interesting that the condensed system appears to be composed of nearly two  dense ``mini-$\alpha$" clusters. Such dense $\bar{K}$-bound states are aptly named  ``$\bar{K}$ cluster."   

\noindent
{\bf Multi-$\bar{K}$ nuclei}\\

\begin{figure}[h]
\begin{center}
\includegraphics[width=0.7\textwidth]{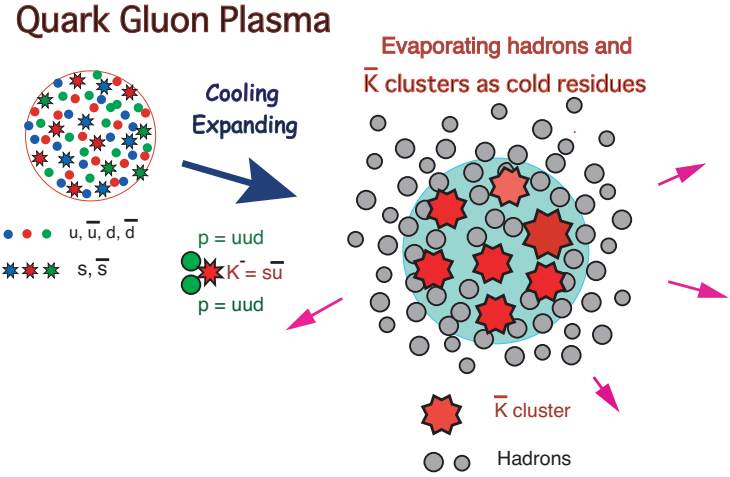}
\end{center}
\vspace{0.5cm}
   \caption{\label{fig:QGP} Quark gluon plasma and its transition to evaporating hadron gases with heavy and dense residues of $\bar{K}$~clusters. }
\end{figure}

Double $\bar{K}$ nuclei, such as $ppK^-K^-$, $ppnK^-K^-$ and $pppK^-K^-$, are predicted to exist  with much larger binding energies, almost twice as large as in single $\bar{K}$ systems \cite{Yamazaki:02,Yamazaki:04}. How can we produce them? In addition to the $(K^-,K^+)$ type reactions (and through their continuum compound reactions), we propose to search for them in heavy-ion reactions at high energy \cite{Yamazaki:04}. First, a cascade evolution of $\bar{K}$ clusters (capture reactions), in which the $\bar{K}$ falls into a state of lower and lower energy by capturing surrounding nucleons, like a self-trapping center, may take place. Secondly,  Fig.~\ref{fig:QGP} shows how a dense $\bar{K}$ cluster formed in a hot plasma remains in a cooling stage, like a microscopic solid residue in a liquid/gas. To identify $\bar{K}$ clusters we employ invariant-mass spectroscopy method, which is applicable when all the decay channels can be measured and the decay occurs after the freeze-out phase.

\noindent
{\bf Why does $\bar{K}$ produce shrunk nuclear systems?}\\

Why are such high-density nuclear states  possible, in apparent violation of the nuclear physics ``law" of {\it constant nuclear density}? The constant density of nuclei is maintained by the hard core part of the $N-N$ interaction, which may result from the Pauli blocking in the $u-d$ quark sector. Usual hadrons are subject to short-range repulsion, because of the $(u, d)$ Pauli blocking. In normal nuclei, the average inter-nucleon distance is $d_{NN} \sim 2.2$ fm. The nucleon rms radius, $r_{\rm rms} \sim 0.86$ fm, corresponds to a nucleon volume of $v_N \sim$ 2.66 fm$^3$ and to a nucleon density of $\rho_N \sim$ 0.38 fm$^{-3}$. This means that nucleons occupy the nuclear space with a compaction factor of $f_c = \rho_N/\rho_0 \sim$ 2.3. This situation is almost unchanged.

The exceptional case is seen in $\bar{K}$, composed of $s\bar{u}$ ($\bar{K}^0$) or $s\bar{d}$ ($K^-$), which includes no $(u,d)$ quark. Thus, the $\bar{K} N$ interaction is dominated by the particle-antiparticle attraction without a hard core. (A similar exceptional situation exists for $D$ mesons). From this consideration, we understand why normal nuclei are difficult to compress and why only the $\bar{K}$ meson mediates dense nuclear systems. The $\bar{K}$ meson is an intruder to relax the $NN$ hard core, thus increasing the average nucleon density to $ \langle \rho \rangle \sim 3 \rho_0$, which exceeds the above nucleon compaction factor at which the QCD vacuum is expected to vanish and chiral symmetry is restored. It is vitally important to investigate to what extent the involved hadrons keep their identities in such an extremely dense system. 
In this respect, $\bar{K}$ clusters can be viewed as particle-antiparticle systems which, however, are not as violent as in $\bar{p}$ nuclear systems. \\
\begin{figure}[t!]
\begin{center}
\includegraphics[width=0.3\textwidth]{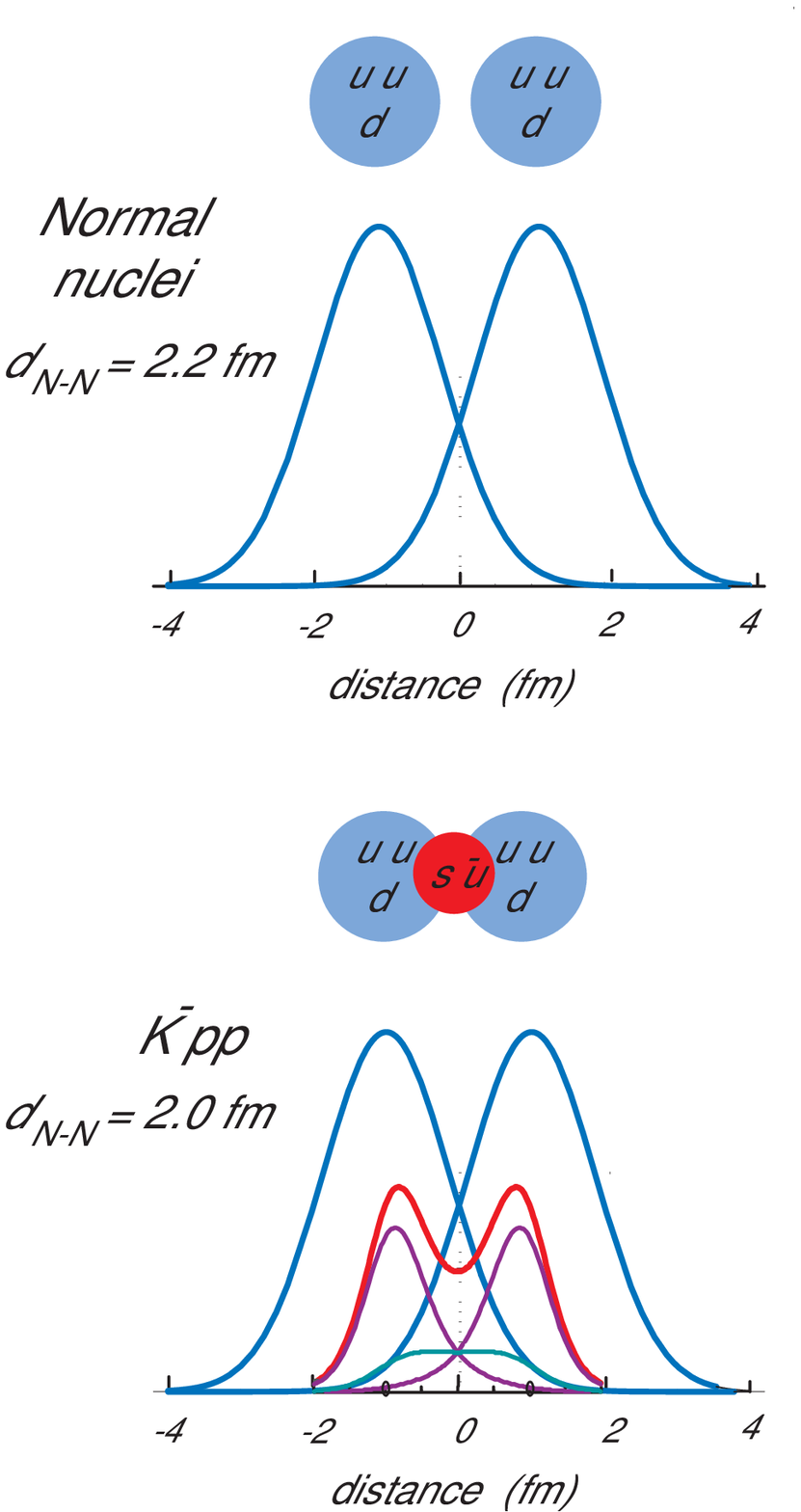}
\includegraphics[width=0.5\textwidth]{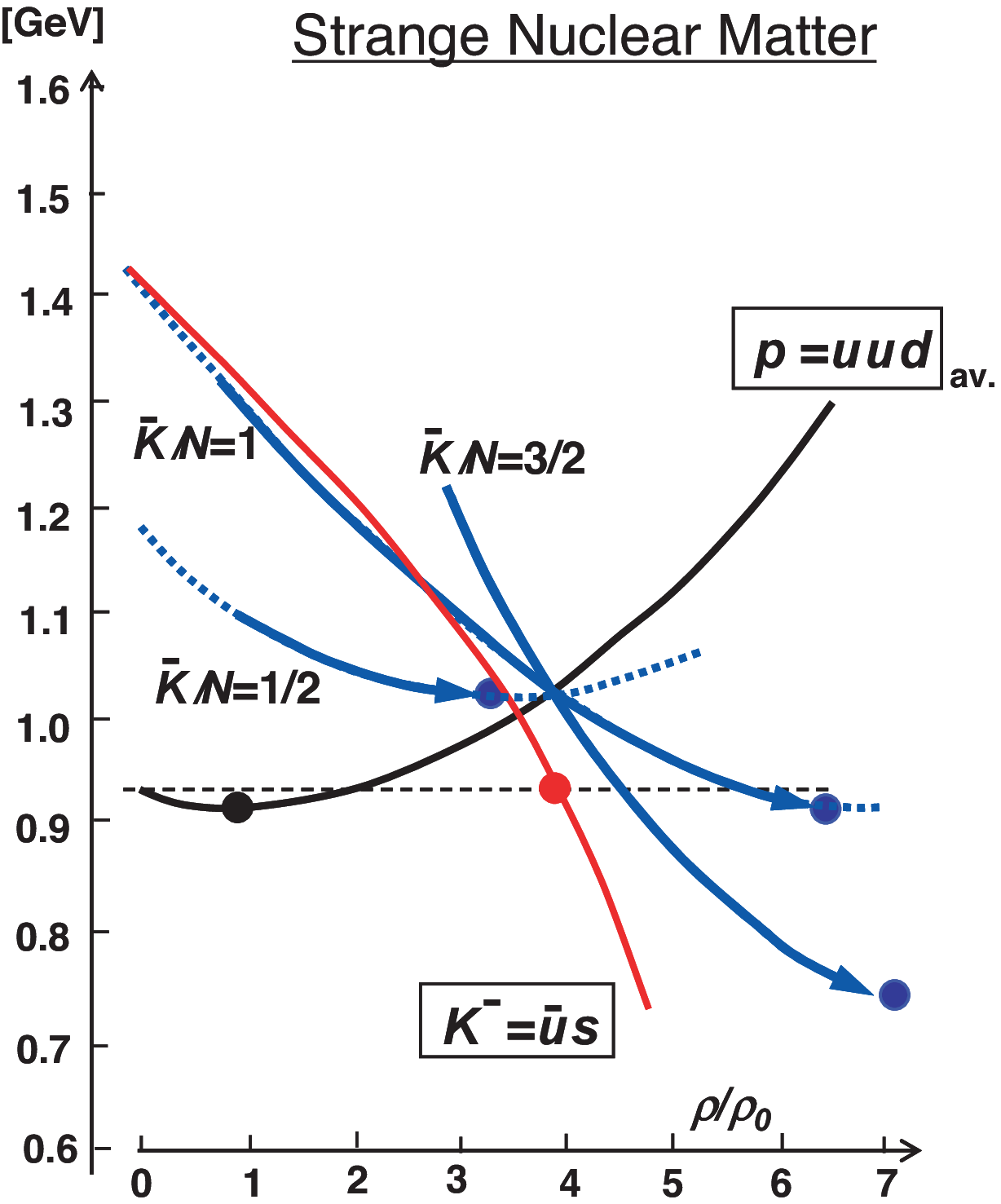}
\end{center}
   \caption{\label{fig:Kmatter} (Left) The density distributions of the protons and $K^-$ in $K^-pp$ in comparison with the normal configuration in ordinary nuclei. (Right) 
    Speculated diagrams for the density dependences of the bound-state energies of various baryon composite systems ($pK^-)^{m}$n$^{n}$. The $\bar{K} N$ energy is represented by the  red line/curve, and the nuclear compression by the black curve. The total energies for representative fractions of $K^-/N$ (=1/2, 1 and 3/2) are depicted by respective blue curves, which possess minima at high density and low energy. The case of density-dependent enhanced $\bar{K}N$.}
\end{figure}

\noindent
{\bf Kaon condensation}\\

The above consideration naturally leads us to a regime of kaon condensation \cite{Kaplan:86,Brown:94}. Specifically, $\bar{K}$ mesons, as intruders with $\bar{u}$ and $\bar{d}$ quarks, behave as a strong glue to combine surrounding nucleons into a dense system. The total energy drops by an amount that depends on the composition of $p$, $n$ and $\bar{K}$. Intuitively, one can construct energy diagrams like Fig.~\ref{fig:Kmatter}. 
Thus, the microscopic $\bar{K}$-bound nuclear clusters, which we are now studying, are building blocks of large $\bar{K}$ matter. Not only does their existence itself have fundamental importance from the viewpoint of dense and bound quark-gluon systems, but also, it is pertinent to the problem of strange matter and stars.
Despite the drastic dynamical change of the system caused by the strong $\bar{K} N$ interaction the identity of the ``constituent atom", $\Lambda^*$, is nearly preserved because of the presence of a short-range repulsion between the two protons. In the same sense, the previously predicted $K^-K^-pp$ \cite{Yamazaki:04} corresponds to the two-electron neutral hydrogen molecule (H$_2$). 
Thus, we have come to {\it super strong nuclear force}, a revival of the Heitler-London-Heisenberg scheme. We can summarize that dense kaonic nuclear clusters are accommodated by this super strong nuclear force without the aid of gravity.

\section*{Acknowledgements}
The author would like to thank deeply his intimate collaborators, Professors H. Toki, S. Hirenzaki, P. Kienle, R.S. Hayano, and Y. Akaishi, for the active collaborative work. One of us (T.Y.) is grateful to the Alexander von Humboldt Foundation for its ``Forschungspreis", and acknowledges the receipt of Grant-in-Aid for Scientific Research of Monbu-Kagakusho of Japan.

\end{document}